\documentclass[sigconf,natbib=true]{acmart}

\renewcommand\footnotetextcopyrightpermission[1]{} 
\AtBeginDocument{%
  \providecommand\BibTeX{{%
    \normalfont B\kern-0.5em{\scshape i\kern-0.25em b}\kern-0.8em\TeX}}}

\setcopyright{acmcopyright}
\copyrightyear{2022}
\acmYear{2022}
\acmDOI{XXXXXXX.XXXXXXX}

\acmConference[Conference acronym 'XX]{Make sure to enter the correct
  conference title from your rights confirmation emai}{June 03--05,
  2018}{Woodstock, NY}
\acmPrice{15.00}
\acmISBN{978-1-4503-XXXX-X/18/06}

\usepackage{algorithm}
\usepackage{algorithmic}
\usepackage{hyphenat}
\usepackage{color}
\usepackage{multirow}

\newcommand{\name}{$\mathtt{EdgeNet}$}
\begin{document}

\title{EdgeNet~:~\underline{E}ncoder-\underline{d}ecoder \underline{ge}nerative \underline{Net}work for Auction Design in E-commerce Online Advertising}

%
\author{Guangyuan Shen}
\authornote{Both authors contributed equally to this research.}
\author{Shenjie Sun}
\authornotemark[1]
\email{{shenguangyuan.sgy,shengjie.ssj}@alibaba-inc.com}
\affiliation{%
  \institution{Alibaba Group}
  \city{Hangzhou}
  \state{Zhejiang}
  \country{China}
}

\author{Dehong Gao}
\authornote{Corresponding Author}
\author{Libin Yang}
\email{{dehong.gdh,libiny}@nwpu.edu.cn}
\affiliation{%
  \institution{NWPU}
  \city{Xian}
  \state{Shaanxi}
  \country{China}
}

\author{Yongping Shi}
\author{Wei Ning}
\email{{yongping.syp, wei.ningw}@alibaba-inc.com}
\affiliation{%
  \institution{Alibaba Group}
  \city{Hangzhou}
  \state{Zhejiang}
  \country{China}
}




\begin{abstract}
We present a new encoder-decoder generative network dubbed EdgeNet, which introduces a novel encoder-decoder framework for data-driven auction design in online e-commerce advertising.
We break the neural auction paradigm of Generalized-Second-Price (GSP) and improve the utilization efficiency of data while ensuring the economic characteristics of the auction mechanism.
Specifically, EdgeNet introduces a transformer-based encoder to better capture the mutual influence among different candidate advertisements.
In contrast to GSP based neural auction model, we design an auto-regressive decoder to better utilize the rich context information in online advertising auctions.
EdgeNet is conceptually simple and easy to extend to the existing end-to-end neural auction framework.
We validate the efficiency of EdgeNet on a wide range of e-commercial advertising auctions, demonstrating its potential in improving user experience and platform revenue.
\end{abstract}




\received{20 February 2007}
\received[revised]{12 March 2009}
\received[accepted]{5 June 2009}

\maketitle

\section{Introduction}
Online e-commerce advertising has grown into a massive industry, both in terms of advertisement(ad) volume as well as the complexity of the mechanism design behind~\cite{feng2019online}. 
Traditional auction mechanisms, such as Vickrey-Clarke–Groves (VCG) auction~\cite{vickrey1961counterspeculation}, Myerson auction~\cite{myerson1981optimal} and generalized second-price auction~\cite{edelman2007internet}, have been used to enable efficient ad allocation in various e-commerce advertising scenarios. 
However, none of these methods can make good use of the rich user history data of online advertising to optimize allocation and payment rules. 
It remains open to both academia and industry on how to make full use of powerful deep learning in designing data-driven auction mechanisms for industrial e-commerce advertising.

Recently, pioneered by Liu et al.~\cite{liu2021neural}, there is rapid progress in designing data-driven auctions through deep learning~\cite{liu2021neural,duetting2019optimal,liao2022nma}.
Typically, we can formulate a data-driven auction design as a constrained optimization problem and find near-optimal solutions~\cite{conitzer2002complexity,conitzer2004self}.
The data-driven auctions enable us to exploit rich information, such as the context of the auction environment and the performance feedback from auction outcomes, to guide the design of a flexible mechanism.
Though effective, most existing data-driven auction frameworks~\cite{liu2021neural,liao2022nma} are still “restricted” to the standard paradigm of Top-K ranking: Map the contextual auction features to one-dimension rank score space, and then perform top-k sorting to generate new allocation and second price payment results.
Such a paradigm may not be optimal in e-commerce online advertising for two main reasons.

\begin{figure}
    \centering
    \includegraphics[width=\linewidth]{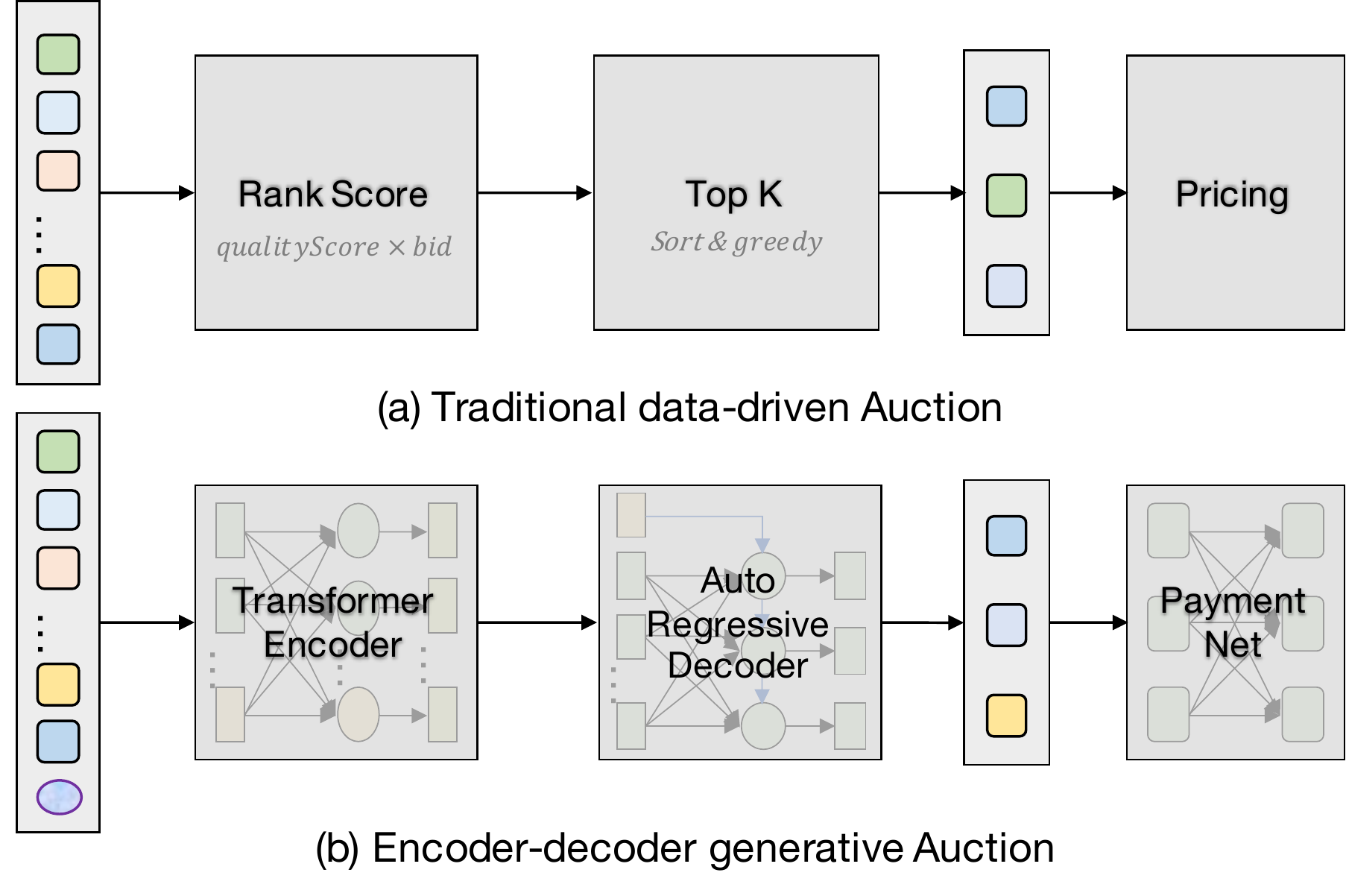}
    \caption{The difference between GSP-based neural auction and encoder-decoder generative auction framework.}
    \label{fig:topk}
\end{figure}
\textbf{First, top-k ranking allocation has limited power to utilize auction contextual information.} According to the greedy sorting method, such auction models can not know the final allocation result of each candidate ad when they output the rank score.
As illustrated in~Fig.\ref{fig:topk}, the contextual information of these candidate ads when they are auctioned is different from the contextual information when they are displayed.
However, all the existing e-commerce neural auction models~\cite{liu2021neural,liao2022nma} still assign a rank score to each ad only based on the original context ignoring the change in the contextual information.
%
At this time, the estimated display revenue (rank score) of the top-ranked ads may not be the highest since there exists mutual influence among the exposed ads.
In practice, if an ad item is surrounded by others with similar quality but much higher prices, then its probability of being clicked would be high. 
On the contrary, if the same item is surrounded by items of much lower prices, then its probability of being purchased would be lower. 
Therefore, in e-commerce advertising auctions, mutual influences between items are even stronger than those in traditional auctions. 

\textbf{Second, second price payment cannot leverage the full power of data-driven auctions.} 
Almost all the data-driven auctions for e-commerce still use the GSP payment rule~\cite{liu2021neural,liao2022nma}, that is, paying the second-highest bid with a rank score fraction discount.
Such a second-price payment paradigm can guarantee the economic characteristics of the mechanism design, like, Incentive Compatible (IC), and Individual Rational (IR).
However, it is impossible to quantify the correct "distance" between two ads in the final impression context since the rank score itself is obtained based on the original auction context.
How to break the GSP paradigm in e-commerce advertising auction design while making the auction conform to economic characteristics is still an open problem for researchers from industrial and academic institutes.
To handle such practical problems, we need a new architecture to better model the auction context while maintaining economic characteristics.

To overcome the aforementioned limitations of the previous works, we propose \name: an \underline{E}ncoder-\underline{d}ecoder \underline{ge}nerative neural \underline{Net}work architecture as the auction model to be optimized. 
The context encoder is built upon the transformer architecture~\cite{vaswani2017attention}, which can capture the complex mutual influence among different ads and user page view information in an auction.
In the auto-regressive context decoder, we generate auction results one by one, i.e., once we have selected a candidate ad we will update the context information immediately and then select the next ad, thus it can perceive the dynamically changing auction context. 
Moreover, instead of following the GSP auction paradigm, we design lightweight ex-post regret loss to approximate the \underline{D}ominant-\underline{S}trategy \underline{I}ncentive \underline{C}ompatibility~(DSIC).
We have deployed the \name mechanism in the real advertising system for the e-commerce platform. 
Experimental results on large-scale industrial data sets showed that \name~mechanism significantly outperformed other widely used industrial auction mechanisms in optimizing multiple performance metrics.
Our main contributions can be summarized as follows:
\begin{itemize}
    \item We are the first to realize that the GSP auction paradigm may not be optimal in e-commerce advertising since it can not model the change of the auction context and ignore the mutual influence among different candidate ads.
    \item We provide the first insight to model the context change and candidate mutual influence in auction design. The proposed \name~casts auction as a sequence generative task and outputs the allocation results one by one, which can fully perceive the context change.
    \item We break the limit of the second-price payment, and output the payment fraction based on the payment network. We present the first ex-post regret loss training task in e-commerce advertising to approach DSIC.
\end{itemize}

\section{Preliminaries}
\label{sec:preliminaries}
\subsection{Data-driven auction design}
Similar to prior work~\cite{liu2021neural,zhou2018deep}, we describe a typical ad platform for online e-commerce.
$N$ advertisers compete for showing their ads in $K \leq N$ ad slots, which are incurred by a page view request from the user.
Each advertiser $i$ submits bid $b_i$ based on his private information including the predicted click-through rate~($pCTR$), predicted conversion rate~($pCVR$), cost per conversion~($CPC$), etc, over the ad.
We use vector $\mathbf{b}=(b_i,\mathbf{b}_{-i})$ to represent the bids of all advertisers, where $\mathbf{b}_{-i}$ are the bids from all advertisers except $i$.
We represent the ad auction mechanism by $\mathcal{M}\langle\mathcal{R},\mathcal{P}\rangle$, where $\mathcal{R}$ is the allocation rule and $\mathcal{P}$ is the payment rule.

A data-driven auction mechanism $\mathcal{M}\langle\mathcal{R},\mathcal{P}\rangle$ consists of an allocation rule $\mathcal{R}$ and a payment rule $\mathcal{P}$:
The allocation rule $\mathcal{R}=(\mathcal{R}_{ij})_{i\in N, j\in K}$ computes the probability that ad slot $j$ is allocated to candidate ad $i$, given the bidding profile $b$, candidate ad information $x$ and user information $y$. 
For all $b, {{x}}, {{y}}$, and $j \in K$, we have $\sum_{i=1}^N \mathcal{R}_{ij}(b, {{x}}, {{y}})\leq 1$ to guarantee no slot is allocated more than once. 
The payment rule $p=(p_1,p_2,\dots,p_n)$ computes the price advertiser $i$ need to pay.
\begin{definition}[Utility]
In a data-driven auction setting, the utility of advertiser $i$ under mechanism $\mathcal{M}\langle\mathcal{R},\mathcal{P}\rangle$ is defined by
\begin{equation*}
    u_i({v}_i, b,{{x}},{{y}})= \sum_{j=1}^K \mathcal{R}_{ij}(b,{{x}},{{y}}) v_{ij} - p_i(b,{{x}},{{y}}),
\end{equation*}
where $v_{ij} = pCTR\times pCVR \times CPC$ denote the valuation of advertiser $i$ wins the ad slot $j$ in the auction.
\end{definition}
The ad allocation rule would jointly consider the bids and the quality ($pCTR$ and $pCVR$) of the ads. 
We use $\mathcal{R}_{ij}(b_i, \mathbf{b}_{-i})=1$ to denote the advertiser $i$ wins the $j^{th}$ ad slot, while $\mathcal{R}_{ij}(b_i, \mathbf{b}_{-i})=0,~\forall j \in K$ represents the advertiser loses the auction.
The $K$ winning ads would be displayed to the user.
The auction mechanism module further calculates the payments for the winning ads with a rule $\mathcal{P}$, which would be carefully designed to guarantee the economic properties and the revenue of the auction mechanism.  
\begin{definition}[DSIC]
\label{def:DSIC}
An auction is \emph{dominant strategy incentive compatible} (DSIC) if for each advertiser, the optimal strategy is to report her true valuation no matter how others report.
\end{definition}
\begin{definition}[IR]
\label{def:IR}
An auction is \emph{individually rational} (IR) if for each advertiser, truthful bidding will receive a non-negative utility.
\end{definition}
\begin{definition}[(Ex-post) Regret]
The ex-post regret for an advertiser $i$ under auction is the maximum utility gain he can achieve by misreporting when the bids of others are fixed, i.e., 
\begin{equation*}
\begin{aligned}
    rgt_{i}(v,{{x}},{{y}}) := \max_{{b}_i \pm~\epsilon} \{u_i({v}_i, ({b}_i\pm\epsilon, b_{-i}),{{x}},{{y}})
    -u_i({v}_i, b,{{x}},{{y}})\}.
\end{aligned}
\end{equation*}
All the expectation terms are computed empirically by $L$ samples, sampling from our train data sets. The empirical ex-post regret for advertiser $i$ is defined as
\begin{equation} 
\label{eq:emp_reg}
\begin{aligned}
\widehat{rgt}_i := \frac{1}{L}\sum_{\ell=1}^L rgt_i(v^{(\ell)}, x^{(\ell)}, y^{(\ell)}), 
\end{aligned}
\end{equation}

\end{definition}
\subsection{Problem Formulation}
Following the work~\cite{zhang2021optimizing,liu2021neural}, we formulate the problem as \emph{multiple performance metrics optimization in the competitive advertising environments}. Given bid vector $\textbf{b}$ from all the advertisers and $L$ ad performance metric functions ${\{f_1(\mathbf{b};\mathcal{M}), .., f_L(\mathbf{b};\mathcal{M})\}}$ (such as Revenue, CTR, CVR, etc), we aim to design an auction mechanism $\mathcal{M}\langle \mathcal{R}, \mathcal{P} \rangle$, such that
\begin{equation}
\begin{aligned}
    \max_{\mathcal{M}} \quad & \mathbb{E}_{\mathbf{b} \sim \mathcal{D}} [F(\mathbf{b};\mathcal{M})]\\
\textrm{s.t.} \quad 
& \textit{Dominant Strategy Incentive Compatible Constraint}\\
\end{aligned}
\label{eq:problem}
\end{equation}
where $\mathcal{D}$ is the advertisers' bid distribution based on which bidding vectors $\mathbf{b}$ are drawn. We define $F(\mathbf{b};\mathcal{M}) = \lambda_1 \times f_1(\mathbf{b};\mathcal{M}) + \cdots + \lambda_L\times f_L(\mathbf{b};\mathcal{M})$, where the objective is to maximize a linear combination of the multiple performance metrics $f_l$'s with preference parameters $\lambda_l$'s. 
The parameters $\lambda_l$'s are the inputs of our problem. The constraints of DSIC guarantee that advertisers would  truthfully report the bid.

\section{Model Architecture}
\label{sec:net}
\begin{figure*}[ht]
  \includegraphics[width=\linewidth]{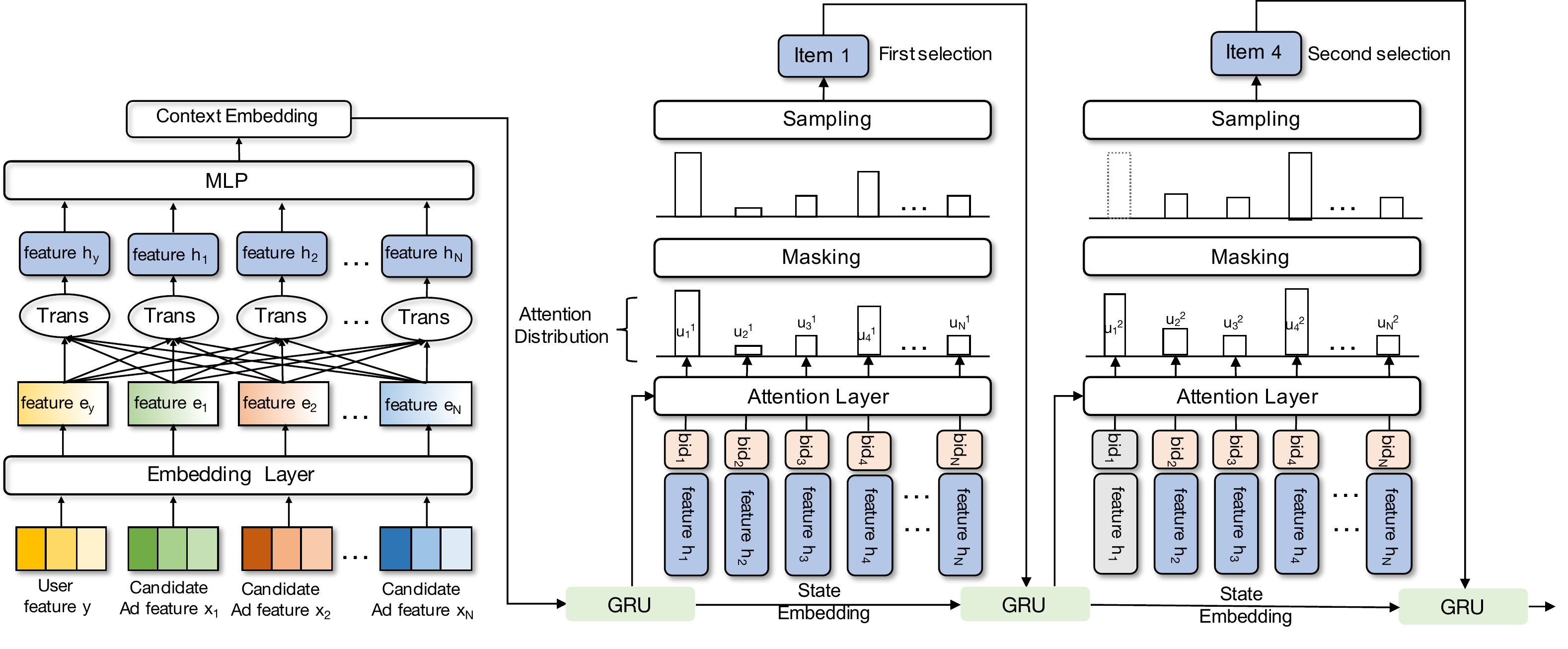}
  \caption{The encoder-decoder generative neural network for auction design. The masking is used to ensure that one advertiser can only win one slot, while the sampling is designed for exploration. }
   \label{fig:encoder_decoder}
\end{figure*}
\subsection{Overview of \name}
As illustrated in Fig.~\ref{fig:encoder_decoder}, \name~consists of two main parts: a permutation-equivariant context encoder and an auto-regressive auction decoder.
The transformer-based context encoder learns an auction context embedding from the features of page view user and candidate ads.
The structural properties of the transformer not only allow for a better representation of the competition among different candidate ads but also guarantee that the context feature is permutation-equivariant.
Afterward, we employ an auto-regressive decoder to generate context-aware exposure ads sequence. 
This neural network is partially monotonic with respect to bids, which is critical to the guarantee of IC property. 
Finally, we compute the allocation and payment result through the final output layer.
\subsection{Auction Context Encoding}
Given the features of each candidate ads $x_{i} \in \mathbb{R}^{d_{x}}$ and user $y \in \mathbb{R}^{d_{y}}$, each instance $\mathbf{x}_i$ and $y$ is firstly mapped to a dense and continuous space through an embedding layer, resulting in a set of intermediate states $\mathbf{e} = \{e_i\}^{N+1}_{i=1}$:
\begin{equation}
    e_{i} = \mathrm{Embedding}(x_{i}) \in \mathbb{R}^{d_{x}^{'}},~e_{y} = \mathrm{Embedding}(y) \in \mathbb{R}^{d_{y}^{'}}
\end{equation}
Then, this intermediate states set $\mathbf{e} = \{e_i\}^{N+1}_{i=1}$ is processed with Transformer~\cite{vaswani2017attention} to build the context-aware embedding $h_{i}$ for each candidate ad $i$ and user embedding $h_{y}$ ,
\begin{equation}
    h_{i} = \mathrm{Transformer}(e_{y},e_{i}, e_{-i})\in \mathbb{R}^{d_{h}},
\end{equation}
where $e_{-i}$ denote the intermediate state set  $\mathbf{e} = \{e_1,...,e_{i-1},e_{i+1},...,e_{N}\}$ except $e_{i}$.
The final context embedding $\mathbf{c}$ can be obtained through an extra fully connected layer $\phi$:
\begin{equation}
    \mathbf{c} = \phi(h_{1}, h_{2},...,h_{N+1})\in \mathbb{R}^{d_{c}}.   
\end{equation}
It should be noted that the context encoder does not include the bids from all candidate ads. 
This design is specified mainly for the guarantee of IC property, keeping the property that the advertiser would win the same or a better ad slot if she reports a higher bid.
\subsection{Auto-regressive Decoding}

To break through the limitation of the GSP auction paradigm and better utilize the rich contextual features of online advertising auctions, we propose a novel auto-regressive decoder.
This decoder can generate auction results one by one thus it can perceive the real and dynamically changing auction context.
Once the decoder selects an ad from all the candidate ads set, it immediately updates the context information and then selects the next exposure ad.
The selection is based on the attention mechanism enhanced by context embedding.
As shown in Fig.~\ref{fig:encoder_decoder}, at the beginning of the decoding, the context embedding will be used as the initial hidden state of the GRU cell~\cite{dey2017gate}, and then a special token "start" will be fed into the GRU cell as the initial input.
After that, at each step, the output embedding of the GRU cell will be used as the state embedding, which should contain all the information needed to select the next ad.
The model will consider the pre-order context information when selecting an ad at each step, and then update this context information to affect subsequent selection.
Formally, we have the logits as follows,
\begin{equation}
    \mu_{i}^{j} = v^{T}tanh(\mathbf{W_1}h_{i}+\mathbf{W_{2}}c_{j})+e^{\mathbf{W_3}}b_{i},
\end{equation}
where $v^{T}, W_1, W_2, W_3$ are the model parameter to be optimized, $h_{i}$ denote the feature embedding of ad $i$ and $c_{j}$ denote the state embedding of slot $j$.
In the last decoder layer, we employ an MLP layer to get the global feature maps, which will be used to compute the final allocation and payment in the output layer.
\begin{table*}[ht]
  \renewcommand\arraystretch{1.2}
  \caption{The experimental results of four methods on two datasets. Each result is presented in the form of mean $\pm$ standard deviation.}
  \setlength{\tabcolsep}{3.3mm}{
  \begin{tabular*}{0.95\textwidth}{l|l|cccc}
  \hline
  Dataset &  Model           & CTR               & RPM              & CVR             & $\boldsymbol{\Psi}$               \\
  \hline 
  \hline
  \multirow{5}{*}{Our Dataset}
  &  GSP               & 0.8929 $\pm$ 0.0011\ (-10.71\%) & 0.8760 $\pm$ 0.0018\ (-12.40\%) & 0.9751 $\pm$ 0.0005\ (-2.49\%) & 4.95\% \\
  & uGSP               & 0.9286 $\pm$ 0.0021\ (-7.14\%) & 0.8528 $\pm$ 0.0021\ (-14.72\%) & 0.9982 $\pm$  0.0009\ (-0.18\%) & 12.19\% \\
  & DNA              & 0.9405 $\pm$ 0.0013\ (-5.95\%) & 0.9070 $\pm$ 0.0029\ (-9.30\%)  & 0.9876 $\pm$ 0.0010\ (-1.24\%) & 3.08\% \\
  & \name       & \textbf{1.0000 $\pm$ 0.0015} & \textbf{1.0000 $\pm$ 0.0020 } & \textbf{1.0000 $\pm$ 0.0008} & \textbf{2.09\%}   \\
  \hline 
 
  \end{tabular*}}

  \label{tab:result}
\end{table*}
\begin{equation}
    F =(F^{\mathcal{R}},F^{\mathcal{P}})= \mathrm{MLP}(\mu_{j}^i), \forall i \in N, \forall j \in K
\end{equation}
The first feature map $F^{\mathcal{R}} \in \mathbb{R}^{N\times K}$ is used to compute the original allocation probability $\mathcal{R}(b,{x},y) \in [0, 1]^{N\times K}$ by softmax activation function on each column of $F^{\mathcal{R}}$, i.e., 
\begin{equation}
    \mathcal{R}_{i,j}= \mathrm{Softmax}(F^{\mathcal{R}}_{\cdot, j}), \forall j \in K.
\end{equation}
Here $\mathcal{R}_{i,j}$ is the probability that slot $j$ is allocated to ad $i$.
For payment, we compute payment fraction $\tilde p(b,{x},y) \in (0, 1)^n$ via the second feature map $F^p$:
\begin{equation}
    \tilde p_i = \mathrm{Sigmoid}\big(\frac{1}{K}\sum_{j=1}^K F^p_{ij}\big), \forall i\in N,
\end{equation}

\subsection{Optimization and Training}
Similar to~\cite{duetting2019optimal}, \name~is optimized through the augmented Lagrangian method.
The Lagrangian with a quadratic penalty is:
\begin{equation}
    \mathcal{L} = -\sum_{i=1}^N\{\sum_{j=1}^{K} \mathcal{R}_{i,:} \cdot F_{all}-\tilde p_i\cdot b_{i}\}+\sum_{i=1}^N\rho_i\widehat{rgt}_i+\frac{\rho}{2}\sum_{i=1}^N\left(\widehat{rgt}_i\right)^2
\end{equation}
where $F_{all}=[\sum_{l=1}^{L}\lambda_l\times f^1_l,\cdots, \sum_{l=1}^{L}\lambda_l\times f^N_l]^T$.
\section{Experimental Evaluation}
\subsection{Experiment Setup}
\subsubsection{Datasets}
The data sets we used for experiments come from an e-commerce advertising system. 
We randomly select 1 million records logged data under GSP auctions from Sept. 1-4, 2022 as training data, and 1200k records logged data from Sept. 5, 2022 as test data. Unless stated otherwise, all experiments are conducted under the setting of top-3 ads displayed in each page view~\footnote{We are now working on data desensitization for public release}. 
\subsubsection{Evaluation Metrics}
We consider the following metrics in our offline experiments, which reflect the platform revenue, user experience in e-commerce advertising. For all experiments in this paper, metrics are normalized to a same scale.
\begin{itemize}
    \item Revenue Per Mille (RPM). $RPM = \frac{\sum click\times PPC}{\sum impression}\times1000$.
    \item Click-Through Rate (CTR). $CTR = \frac{\sum click}{\sum impression}$. 
    \item Conversion Rate (CVR). $CVR = \frac{\sum order}{\sum impression}$.
    \item Incentive Compatiable-Regret (IC-R)~\cite{wang2022designing}, which represents the exposed regret of utility maximizers, to quantify IC of \name. A larger value of IC-R indicates that an advertiser could get larger utility by manipulating the bidding. For instance, 2.09$\%$ in Table~\ref{tab:result} means advertisers can increase their utilities by about 2.09$\%$ through modifying bid in \name~auctions.
\end{itemize}

\subsubsection{Baselines Methods} We compare \name~with the widely used mechanisms in the industrial ad platform.

\textbf{1) Generalized Second Price auction (GSP).} The rank score of traditional GSP is simply the bids times $pCTR$, that is, effective Cost Per Milles (eCPM). The payment rule is the value of the minimum bid required to retain the same slot. The work~\cite{lahaie2007revenue} suggested incorporating a squashing exponent $\sigma$ into the rank score function, i.e., $bid \times pCTR^{\sigma}$ could improve the performance, where $\sigma$ can be adjusted to weight the performance of revenue and CTR. We refer to this exponential form extension as GSP in the experiments.

\textbf{2) Utility-based Generalized Second Price auction (uGSP).} uGSP~\cite{bachrach2014optimising} extends the conventional GSP by taking the rank score as a linear combination of multiple performance metrics using estimated values: $r_i(b_i) = \lambda_1\times b_i\times pCTR_i + o_i$, where $o_i$ represents other utilities, such as pCTR and pCVR: $o_i=\lambda_2\times pCTR_i + \lambda_3\times pCVR_i (\mbox{where }\lambda_l \geq 0)$. The payment of uGSP follows the principle from GSP: $p_i = \frac{\lambda_1\times b_{i+1}\times pCTR_{i+1} +o_{i+1} - o_{i}}{\lambda_1 \times pCTR_i}$. 

\textbf{3) DNA~\cite{zhang2021optimizing}} DNA uses a deep neural network to map ad's related features to a new rank score within the GSP auction.

\subsection{Results}
We construct an offline advertising simulation system.
This simulation system can ensure that the offline and online performance
trends are consistent.
Each experiment is repeated 10 times with different random seeds and each result is presented in the form of mean ± standard. 
We summarize the detailed experimental results on our industrial datasets in Table~\ref{tab:result}. 
Compared with representative auction mechanisms, we have the following observations from the experimental results: 1)~\name~makes improvements over DNA, uGSP, GSP in CTR, RPM, CVR. One reasonable explanation is that \name~have better capability to model the mutual influence among different candidate ad. 2) Compared with traditional GSP-based auction design (e.g., uGSP, DNA), \name~also has a relatively low IC-R, i.e., maintain the economic properties without the help of GSP auction. The regret loss combined with the payment neural network plays an important role in keeping IC.

%


\balance
\section{Conclusion}
We present a novel encoder-decoder generative network (EdgeNet), which introduces a novel encoder-decoder framework for data-driven auction design in online e-commerce advertising.
\name~have broken the restriction of GSP auction and realized a real data-efficient auction structure design.
\name~mechanism significantly outperformed other widely used industrial auction mechanisms in optimizing multiple performance metrics.
For future work, we are interested in how to construct budget-aware data-driven auctions for online e-commerce advertising.

\bibliographystyle{ACM-Reference-Format}
\bibliography{sample-base}

\end{document}